\documentclass[showpacs,aps,prstab,amsmath,amssymb]{revtex4-1}
\usepackage[latin9]{inputenc}
\usepackage{color}
\usepackage{amsmath}
\usepackage{graphicx}
\usepackage{esint}

\makeatletter

\InputIfFileExists{t2aenc.def}{}{%
  \errmessage{File `t2aenc.def' not found: Cyrillic script not supported}}
\DeclareRobustCommand{\cyrtext}{%
  \fontencoding{T2A}\selectfont\def\encodingdefault{T2A}}
\DeclareRobustCommand{\textcyr}[1]{\leavevmode{\cyrtext #1}}

\DeclareTextSymbolDefault{\textquotedbl}{T1}

\usepackage{bm}
\usepackage{booktabs}
\usepackage{color}
\usepackage{hyphenat}

\renewcommand{\[}{\begin{equation}}
\renewcommand{\]}{\end{equation}}
\makeatother

\begin{document}
\title{\textcolor{black}{A model of wave function collapse in a quantum measurement
of spin as the Schr\textcyr{\"\cyro}dinger equation solution of a
system with a simple harmonic oscillator in a bath}}
\author{Li Hua Yu}
\affiliation{Brookhaven National Laboratory}
\date{4/6/2023}
\begin{abstract}
We present a set of exact system solutions to a model we developed
to study wave function collapse in the quantum spin measurement process.
Specifically, we calculated the wave function evolution for a simple
harmonic oscillator of spin $\frac{1}{2}$, with its magnetic moment
in interaction with a magnetic field, coupled to an environment that
is a bath of harmonic oscillators. The system's time evolution is
described by the direct product of two independent Hilbert spaces:
one that is defined by an effective Hamiltonian, which represents
a damped simple harmonic oscillator with its potential well divided
into two, based on the spin and the other that represents the effect
of the bath, i.e., the Brownian motion. The initial states of this
set of wave functions form an orthonormal basis, defined as the eigenstates
of the system. If the system is initially in one of these states,
the final result is predetermined, i.e., the measurement is deterministic.
If the bath is initially in the ground state,and the wave function
is initially a wave packet at the origin, it collapses into one of
the two potential wells depending on the initial spin. If the initial
spin is a vector in the Bloch sphere not parallel to the magnetic
field, the final distribution among the two potential wells is given
by the Born rule applied to the initial spin state with the well-known
ground state width. Hence, the result is also predetermined. We discuss
its implications to the Bell theorem\citep{bell_theorem}.

We end with a summary of the implications for the understanding of
the statistical interpretation of quantum mechanics.
\end{abstract}
\maketitle

\section*{Introduction}

The quantum measurement problem is related to the foundations of quantum
mechanics. In particular, the collapse of the wave function during
quantum measurement is crucial to the understanding of quantum mechanics.
''The inability to observe such a collapse directly has given rise
to different interpretations of quantum mechanics''\citep{measurement}.
We have chosen to use a model for the quantum measurement process
to clarify the collapse of the wave function.

To study the collapse of the wave function, we are interested in the
simplest example of a dissipative system. Such a system, i.e., a harmonic
oscillator coupled to an environment, which is a bath of harmonic
oscillators, has been the subject of extensive studies \citep{Legget1,Legget2,caldeira,kanai,Nakajima,Zwanzig,Senitzky,feynman,kac,kostin,louisell,scully,yasue,koch,dekker,kac2,weisskopf}.
A special case of this system, the Ohmic case discussed by Caldeira
and Leggett (to be defined later), is exactly solved using the path-integral
with the density matrix diagonalized in \citep{Legget1,Legget2,caldeira},
which indicates a wave function collapse. Following these works, we
solved the wave function evolution directly in this special case as
the solution of the Schr\textcyr{\"\cyro}dinger equation\citep{yu},
where even though we did not focus on the description of the wave
function collapse, it is obvious in the solution.

In this paper, for the Ohmic case of the system, we shall focus on
the wave function collapse and introduce spin $\frac{1}{2}$ into
the simple harmonic oscillator, with its magnetic moment interacting
with a magnetic field aligned to the axis of the oscillation in a
direction $z$. The bath of harmonic oscillators serves as a model
of a detector. The potential well of the simple harmonic oscillator
is split into two potential wells according to whether the spin is
in $+z$ or $-z$ direction. Thus, this models a spin measurement
process.

We consider the problem the same as discussed by Caldeira and Leggett
(CL) \citep{Legget1} and \citep{yu}, except that in this case the
main oscillator has spin: a harmonic-oscillator system (the dissipative
system) with coordinate $q$ on $z$ axis, mass $M$, and frequency
$(\omega_{0}^{2}+\Delta\omega^{2})^{\frac{1}{2}}$ , interacting with
a bath of $N$ harmonic oscillators of coordinates $x_{j}$, mass
$m_{j}$, and frequency $\omega_{j}$, where $\Delta\omega^{2}$ is
a shift induced by the coupling already discussed by CL. The Hamiltonian
of the system and the bath is

\begin{align*}
 & H=\frac{p^{2}}{2M}\text{+\ensuremath{\frac{1}{2}M(\omega_{0}^{2}+\Delta\omega^{2})q^{2}-MB\sigma_{z}q+q\sum_{j=1}^{N}c_{j}x_{j}+\sum_{j=1}^{N}\left(\frac{p_{j}^{2}}{2m_{j}}+\frac{1}{2}m_{j}\omega_{j}^{2}x_{j}^{2}\right)} }\\
 & \sigma_{z}\left|+\right\rangle =\left|+\right\rangle =\left(\begin{array}{c}
1\\
0
\end{array}\right),\sigma_{z}\left|-\right\rangle =-\left|-\right\rangle =-\left(\begin{array}{c}
0\\
1
\end{array}\right),\psi=\left(\begin{array}{c}
\psi_{+}(q)\\
\psi_{-}(q)
\end{array}\right)
\end{align*}

The interaction of the spin $\frac{1}{2}\sigma$ dipole moment with
the magnetic field contributes to the term $-MB\sigma_{z}q$, where
z is the projection of the Pauli matrix $\sigma$ on $z$ and $\psi$
is the main oscillator's wave function. The simple harmonic oscillator's
potential well is divided into two potential wells based on the spin.

Here, we outline the solution to this problem with examples and a
possible interpretation:

In Section 1, despite the addition of the new term $-MB\sigma_{z}q$
to the previous Hamiltonian in \citep{yu}, we show that this system
can be solved almost exactly the same way. The system\textquoteright s
time evolution is described by the direct product of two independent
Hilbert spaces: one that is defined by an effective Hamiltonian, representing
a damped simple harmonic oscillator with its potential well divided
into two based on the spin, and the other that represents the effect
of the bath, i.e., the Brownian motion.

We present a set of exact solutions of the system that all collapse
to a $\delta$-function centered at a specified position and time
depending on whether the spin eigenstate is parallel or opposite to
the magnetic field, with a small displacement determined by the eigenstates
of each bath oscillators. The initial states of this set of wave functions
form an orthonormal basis in the Hilbert space of the system, which
is defined as the ``eigenstates of the system.'' Hence, if the system
is initially in one of these states, the final result is predetermined,
i.e., the measurement is deterministic.

In Section 2, if the system is initially a superposition of these
states, the final distribution is a superposition of $\delta$-functions
for a sufficiently long time after the damping time. Each has a probability
amplitude determined by the corresponding eigenstate's probability
amplitude in the initial state at time $t=0$ , with the center position
determined by the spin eigenstate of the main oscillator and the contribution
from each bath oscillator $x_{j}$. Once the distribution is determined
from the initial state by the Born rule, the final distribution is
also predetermined. As a result, we only need to apply quantum mechanics'
statistical interpretation to the initial state; the final distribution
is already determined by the Schr\textcyr{\"\cyro}dinger equation.

One of the examples we shall discuss is the special case where the
initial states of the bath oscillators are in their ground states
at absolute temperature zero. In this case, the initial state of the
bath is not an ``eigenstate of the system,'' as we defined it above,
but a superposition of the ``eigenstates of the system,'' even though
it is in the ground state (we remark here that the system is initially
not in equilibrium even though the bath is in the ground state). The
main harmonic oscillator is assumed to be initially a wave packet
at the origin, i.e., in the middle between the two split potential
wells. If the initial spin state is one of the two eigenstates of
the main oscillator, for example, $\left|+\right\rangle $, the final
state would be at the bottom of one of the two potential wells, that
is, the one with $z>0$, with a narrow spread due to the contributions
from the bath oscillators which is the Brownian width. We find the
width approximately agrees with the well-known width of the ground
state for the simple harmonic oscillator if the damping time is much
longer than its period.

In Section 3, we discuss another example where the initial spin state
is a vector in the Bloch sphere not parallel to the magnetic field,
i.e., it is the eigenstate of $\sigma_{\theta}$ where $\theta$ is
the angle between the vector and the $z$ direction. In this case,
even though the initial state is a pure state, it is not the eigenstate
of the system; it is a superposition of two eigenstates of the system.
This state is not an eigenstate of the spin measurement of $\sigma_{z}$
because the magnetic field is not in the $\theta$ direction. The
probability of the two eigenstates is given by the Born rule applied
to the initial state, i.e., $\cos^{2}(\frac{\theta}{2})$ for $z>0$,
$\sin^{2}(\frac{\theta}{2})$ for $z<0$, in spite of the initial
state being a pure state. We shall demonstrate that the final distribution
has two distinct peaks at the centers of the two potential wells accordingly,
as shown in Fig.1. Again, the final result is determined by the Schr\textcyr{\"\cyro}dinger
equation according to the initial probability distribution.

If the bath temperature is not zero, the system is in a mixed state
specified by a density matrix $\rho$ according to the Boltzmann distribution.
Still, we found the result is similarly determined by initial distribution.

In Section 4, we summarize the result and its implications in the
Bell theorem. Because the discussion is only for a specific model,
the interpretation is only suggestive. Applying the results of Sections
1, 2, and 3, we establish that the analysis of two separate spin measurements
$\sigma_{1}\centerdot\vec{a}$ and $\sigma_{2}\centerdot\vec{b}$
at remote positions A and B on two entangled particles in the Bell
theorem is equivalent to the analysis of two measurements $-\sigma_{2}\centerdot\vec{a}$
and $\sigma_{2}\centerdot\vec{b}$ at time $t=0$ on particle 2 at
the origin.

Because $\sigma_{2}\centerdot\vec{a}$ and $\sigma_{2}\centerdot\vec{b}$
at time $t=0$ do not form a set of commuting observables, if $\sigma_{2}\centerdot\vec{a}$
is definite, then $\sigma_{2}\centerdot\vec{b}$ is uncertain. Hence
the probability distribution is a consequence of the uncertainty principle.
So long as the uncertainty principle is valid, there is no hidden
variable theory to have a definite value for $\sigma_{2}\centerdot\vec{a}$
and $\sigma_{2}\centerdot\vec{b}$ simultaneously. Therefore, there
is no need to resort to ``local hidden variable theory'' as J. Bell
did to prove Bell inequality. As a result, there is no ``nonlocality''
involved in the violation of Bell inequality when the quantum mechanics
prediction is confirmed. Because the result is predetermined by the
initial condition, the final distance of the two entangled particles
in the experiment in the Bell theorem does not affect the probability
distribution as long as their spin correlation is not destroyed by
the environment before detection. Thus, the experiment's agreement
with quantum mechanics does not provide information related to non-locality.
This result suggests a viewpoint that may aid in understanding the
interpretation of Bell's theorem. This result also implies that the
Born rule at the end of the measurement can be derived from the Schr\textcyr{\"\cyro}dinger
equation as long as the Born rule is applied to the interpretation
of the initial state. 

Finally, Section 5 gives a summary of the implications for the understanding
of the statistical interpretation of quantum mechanics.

\section*{1. Solution of a damped simple harmonic oscillator with spin in a
bath}

The dynamic equation for operators in the Heisenberg representation
leads to the following set of equations of motion:

\begin{align}
 & M\ddot{q}=-M\omega_{0}^{2}q-M\Delta\omega^{2}q+MB\sigma_{z}-\sum_{j}c_{j}x_{j}\label{original_eq-1}\\
 & m_{j}\ddot{x_{j}}=-m_{j}\omega_{j}^{2}x_{j}-c_{j}q\ (j=1,2,...,N)\nonumber 
\end{align}

Now, applying the Laplace transform \citep{caldeira} (the bars are
our notations for the Laplace transform, and $s$ is the Laplace transform
of time $t$), Eqs. (\ref{original_eq-1}) can be used to eliminate
the bath variables $x_{j}$ to obtain the equation for $q$,

\begin{align}
\begin{split}\end{split}
 & M(s^{2}\overline{q}-sq_{0}-\dot{q}_{0})=-M\omega_{0}^{2}\overline{q}-M\Delta\omega^{2}\overline{q}+MB\sigma_{z}\frac{1}{s}-\sum_{j}c_{j}\frac{sx_{j0}+\dot{x}_{j0}}{s^{2}+\omega_{j}^{2}}+\sum_{j}\frac{c_{j}^{2}}{m_{j}(s^{2}+\omega_{j}^{2})}\overline{q}\label{laplace eq.-1}
\end{align}

where $q_{0},\dot{q}_{0}$ , $x_{j0},\dot{x}_{j0}$ are the initial
values of the respective operators in the Heisenberg representation.
Assuming the number of bath oscillators is large enough so that we
can replace the sum over j by integration over $\omega_{j}$, the
coefficient of the last term can then be separated into two terms:

\begin{align}
 & \int_{0}^{\omega_{\text{cufoff}}}\frac{c_{j}^{2}}{m_{j}}\frac{\rho(\omega_{j})}{\omega_{j}^{2}}d\omega_{j}-s^{2}\int_{0}^{\omega_{\text{cufoff}}}\frac{c_{j}^{2}}{m_{j}}\frac{\rho(\omega_{j})}{\omega_{j}^{2}}\frac{1}{(s^{2}+\omega_{j}^{2})}d\omega_{j}\label{two_term}
\end{align}

where $\rho(\omega_{j})$ is the bath oscillator density, with an
upper-frequency limit $\omega_{\text{cufoff}}$. Following an argument
similar to the one pointed out by CL \citep{Legget1}, the requirement
that the system becomes a damped oscillator with frequency $\omega_{0}$
and a damping rate $\eta$ in the classical limit, known as the \textquotedbl Ohmic
friction\textquotedbl{} condition, leads to the following constraint:

\begin{equation}
\rho(\omega_{j})=\frac{2\eta M}{\pi}\frac{m_{j}\omega_{j}^{2}}{c_{j}^{2}}\label{rho_def}
\end{equation}

If the frequency renormalization constant $\Delta\omega^{2}$ is chosen
to satisfy

\[
M\Delta\omega^{2}=\int_{0}^{\omega_{\text{cufoff}}}\frac{c_{j}^{2}}{m_{j}}\frac{\rho(\omega_{j})}{\omega_{j}^{2}}d\omega_{j}=\frac{2\eta M}{\pi}\omega_{\text{cufoff}}
\]
Then by observing Eqs.(\ref{two_term},\ref{rho_def}), it can be
shown that for sufficiently large $\omega_{\text{cufoff}}\gg\omega_{0}$,
the first term of Eq.(\ref{two_term}) represents a frequency shift
$\frac{2\eta M}{\pi}\omega_{\text{cufoff}}$, while the second term
of Eq.(\ref{two_term}) leads to a damping term $-\eta Ms\overline{q}$,
with the damping constant $\eta$.

The frequency is shifted to $\omega_{0}$. Then Eq.(\ref{laplace eq.-1})
is simplified, and its inverse Laplace transform yields the quantum
Langevin equation, which is valid at time $t>0_{+}$:

\begin{equation}
\ddot{q}(t)+\eta\dot{q}(t)+\omega_{0}^{2}q(t)=f(t)\label{dampin_eq}
\end{equation}

with a constant magnetic force $B\sigma_{z}$ and the Brownian motion
driving force

\begin{equation}
f(t)=B\sigma_{z}-\sum_{j}\frac{c_{j}}{M}\left(\text{\ensuremath{x_{j0}}}\cos(\omega_{j}t)+\text{\ensuremath{\dot{x}_{j0}}}\frac{\sin(\omega_{j}t)}{\omega_{j}}\right)\label{force}
\end{equation}

During the derivation, in order to carry out the integral in Eq.(\ref{laplace eq.-1}),
we used the requirement of the inverse Laplace transform that $s$
must pass all the singular points from the right side of the complex
plane, and hence Re($s$) \textgreater{} 0. Equations (\ref{dampin_eq})
and (\ref{force}) are the equations of a driven damped harmonic oscillator
with a constant force, the solution of which is well known as a linear
combination of the initial values at $q_{0}-\sigma_{z}d,\dot{q}_{0}$
, $x_{j0},\dot{x}_{j0}$ and a displacement $\sigma_{z}d$, where
$d\equiv\frac{B}{\omega_{0}^{2}}$:

\begin{align}
 & q(t)=a_{1}(t)\left(q_{0}-\sigma_{z}d\right)+a_{2}(t)\dot{q_{0}}+\sigma_{z}d+\sum_{j}\left(\text{\ensuremath{x_{j0}}}b_{j1}(t)+\text{\ensuremath{\dot{x}_{j0}}}b_{j2}(t)\right)\label{qxsol-1}\\
 & x_{i}(t)=\alpha_{i1}(t)\left(q_{0}-\sigma_{z}d\right)+\alpha_{i0}(t)\sigma_{z}d+\alpha_{i2}(t)\dot{q}_{0}+\sum_{j}\left(x_{j0}\beta_{ij1}(t)+\dot{x}_{j0}\beta_{ij2}(t)\right)\nonumber 
\end{align}

with $a_{1}(t)=\frac{\mu e^{-\nu t}-\nu e^{-\mu t}}{\mu-\nu}$ , $a_{2}(t)=-\frac{e^{-\mu t}-e^{-\nu t}}{\mu-\nu}$,
$\mu\equiv\frac{\eta}{2}+\omega i$, $\nu\equiv\frac{\eta}{2}-\omega i$
and here $\omega\equiv\left(\omega_{0}^{2}-\frac{\eta^{2}}{4}\right)^{1/2}$is
the frequency shifted by damping. And $a_{1}(0)=1,a_{2}(0)=0,\dot{a}_{1}(0)=0,\dot{a}_{2}(0)=1$,
(see Appendix I for $a_{1}(t),a_{2}(t),b_{j1}(t),b_{j2}(t)$, etc).
All formulas are correct whether $\omega$ is real or imaginary. We
have redefined the initial time as $t=0_{+}$ to avoid a minor detail
of the initial-value problem. The explicit expressions for $\alpha_{i0}(t),\alpha_{i1}(t),\alpha_{i2}(t),\beta_{ij1}(t),\beta_{ij2}(t)$
are well known in basic physics.

We emphasize that the use of the Laplace transform instead of the
Fourier transform allows us to express $q(t)$ and $x_{j}(t)$ explicitly
in terms of the initial values, as in Eqs. (\ref{qxsol-1}). Equations
(\ref{qxsol-1}) serve as the starting point of subsequent discussions.
We will proceed to find the Green's function of the full system, and
hence the solution of the wave function in the Schr\textcyr{\"\cyro}dinger
representation. The result tells us in what sense the damped oscillator
is described by an effective Hamiltonian without the bath variables
and gives it a specific form. It also shows that under this condition,
the wave function can be factorized, and that the main factor relevant
to the damped oscillator is a solution of the Schr\textcyr{\"\cyro}dinger
equation with an effective Hamiltonian.

Equations (\ref{qxsol-1}) are correct both in classical mechanics
and in quantum mechanics in the Heisenberg representation. We notice
that $q(t)$ and $x_{j}(t)$ are both linear superpositions of $q_{0}-\sigma_{z}d,\dot{q}_{0}$
, $x_{j0},\dot{x}_{j0}$, $\sigma_{z}d$ with c-number coefficients.
The commutation rules between $q(t),\dot{q}(t),x_{j}(t),\dot{x}_{j}(t)$
are $[q(t),\dot{q}(t)]=\frac{i\hbar}{M}~$, $[x_{j}(t),\dot{x}_{j}(t)]=\frac{i\hbar}{m_{j}}$
, and operators $q(t)$ and $\dot{q}(t)$ commute with $x_{j}(t),\dot{x}_{j}(t)$.
One can prove these commutation rules in two ways: (1) by direct computation,
using the fact that at $t=0$, they are correct, and they all commute
with operator $\sigma_{z}$ and (2) by the general principle that
$q(t),\dot{q}(t),x_{j}(t),\dot{x}_{j}(t)$ are related by a unitary
transformation to $q_{0},\dot{q}_{0},\sigma_{z},x_{j0},\dot{x}_{j0}$.

Equations (\ref{qxsol-1}) show that the operators $q(t)$ and $x_{j}(t)$
can each be written as a sum of two terms:

\begin{align}
 & q(t)=\left(Q(t)+\sigma_{z}d\right)+\sum_{j}\xi_{j}(t)\label{Qxi_def-1}\\
 & x_{i}(t)=\zeta_{j}(t)+\sum_{j}X_{ij}(t)\nonumber 
\end{align}

where

\begin{align}
 & Q(t)=a_{1}(t)\left(q_{0}-\sigma_{z}d\right)+a_{2}(t)\dot{q}_{0}=a_{1}(t)Q_{0}+a_{2}(t)\dot{Q}_{0}\label{Qcd}\\
 & \xi_{j}(t)=\text{\ensuremath{x_{j0}}}b_{j1}(t)+\text{\ensuremath{\dot{x}_{j0}}}b_{j2}(t)\nonumber 
\end{align}
$\left(Q(t)+\sigma_{z}d\right)$ and $\zeta(t)$ are linear in $q_{0},\dot{q}_{0},$
and $\sigma_{z}$ and independent of $x_{j0},\dot{x}_{j0}$, and $\xi_{j}(t)$
and $X_{ij}(t)$ are linear in $x_{j0},\dot{x}_{j0}$, and independent
of $q_{0},\dot{q}_{0},$ and $\sigma_{z}$. Thus $Q(t),\sigma_{z}d$
and $\zeta(t)$ are operators in one Hilbert space $S_{Q}$, while
$\xi_{j}(t)$ and $X_{ij}(t)$ are in an independent Hilbert space
$S_{X}$, and the full Hilbert space is a direct \textcolor{black}{product
of $S_{Q}\otimes S_{X}$.}

We shall first analyze the structure of $S_{Q}$. To explicitly show
that we are discussing the $S_{Q}$ space, we define $Q_{0}\equiv q_{0}-\sigma_{z}d$,
$\dot{Q}_{0}\equiv\dot{q}_{0}=-\frac{i\hbar}{M}\frac{\partial}{\partial q_{0}}$.
Because $\sigma_{z}d$ is a constant operator and it commutes with
$q_{0}$ and $\dot{q}_{0}$, we have $[Q_{0},\dot{Q}_{0}]=[q_{0},\dot{q}_{0}]=\frac{i\hbar}{M}$.
Thus, we can write $-\frac{i\hbar}{M}\frac{\partial}{\partial Q_{0}}\equiv\dot{Q}_{0}=\dot{q}_{0}$
as a well-defined operator in the space of $q_{0},\sigma_{z}$, i.e.,
$S_{Q}$. The eigenfunction of $Q(t)$ with an eigenvalue denoted
by $Q_{1}$, in the $Q_{0}$ representation, is easily calculated
to be (see Appendix II),

\begin{equation}
u_{Q_{1}}(Q_{0},t)=\left(\frac{M\omega e^{\frac{\eta}{2}t}}{2\pi\hbar\sin(\omega t)}\right)^{\frac{1}{2}}\exp\left[-i\frac{M}{2\hbar a_{2}}\left(a_{1}Q_{0}^{2}-2Q_{1}Q_{0}+\phi(Q_{1},t)\right)\right]\label{uQ}
\end{equation}

with $\phi$ as an arbitrary phase, i.e., a real number. This eigenfunction
is related to Green's function 

\[
G(Q_{1},Q_{0};t,0)=<Q_{1}|U(t)|Q_{0}>,
\]

where we denote the evolution operator by $U(t$). To see this, we
use the relation between Schr\textcyr{\"\cyro}dinger operator $\mathbf{Q}_{S}$
and Heisenberg operator $Q(t)\equiv\mathbf{Q}_{H}(t)=U^{-1}(t)\mathbf{Q}_{S}U(t)$.

Let $|Q_{1}>$ be the eigenvector of $\mathbf{Q}_{S}$ with eigenvalue
$Q_{1}$, i.e.,

\begin{align*}
 & \mathbf{Q}_{S}|Q_{1}>=Q_{1}|Q_{1}>
\end{align*}

we see that $U^{-1}|Q_{1}>$ is the eigenvector of $Q(t)$ of value
$Q_{1}$

\begin{align*}
 & Q(t)U^{-1}|Q_{1}>=\mathbf{Q}_{H}U^{-1}|Q_{1}>=U^{-1}\mathbf{Q}_{S}UU^{-1}|Q_{1}>=U^{-1}\mathbf{Q}_{S}|Q_{1}>=Q_{1}U^{-1}|Q_{1}>
\end{align*}

So $U^{-1}|Q_{1}>$ is proportional to $u_{Q_{1}}(Q_{0},t)$. If we
choose $u_{Q_{1}}(Q_{0},t)$ as an orthonormal basis, then $U$ is
unitary: $U^{\dagger}=U^{-1}$.

Then, we have $u_{Q_{1}}(Q_{0},t)=<Q_{0}|U^{-1}|Q_{1}>=<Q_{0}|U^{\dagger}|Q_{1}>=<Q_{1}|U(t)|Q_{0}>^{*}=G^{*}(Q_{1},Q_{0};t,0)$,
i.e., $G(Q_{1},Q_{0};t,0)=u_{Q_{1}}^{*}(Q_{0},t)$. Thus we have

\begin{equation}
G(Q_{1},Q_{0};t,0)=\left(\frac{M\omega e^{\frac{\eta}{2}t}}{2\pi\hbar\sin(\omega t)}\right)^{\frac{1}{2}}\exp\left[i\frac{M}{2\hbar a_{2}}\left(a_{1}Q_{0}^{2}-2Q_{1}Q_{0}+\phi(Q_{1},t)\right)\right]\label{greens_function}
\end{equation}

Next, we shall determine the arbitrary phase $\phi(Q_{1},t)$ , which
is the phase of the eigenvectors of $Q(t)$. Using Eq.(\ref{Qcd}),
we find the commutation rule for $Q$ and $\dot{Q}$ (see Appendix
I):

\[
[Q,\dot{Q}]=\left(a_{1}(t)\dot{a_{2}}(t)-\dot{a}_{1}(t)a_{2}(t)\right)[Q_{0},\dot{Q_{0}}]=e^{-\eta t}[Q_{0},\dot{Q_{0}}]=e^{-\eta t}\frac{i\hbar}{M}
\]

Thus, we define the canonical momentum $P(t)$ as

\begin{equation}
P=Me^{\eta t}\dot{Q}=Me^{\eta t}\left(\dot{a}_{1}(t)Q_{0}+\dot{a}_{2}(t)\dot{Q}_{0}\right)=Me^{\eta t}\dot{a}_{1}(t)Q_{0}-i\hbar e^{\eta t}\dot{a}_{2}(t)\frac{\partial}{\partial Q_{0}}\label{Pdef}
\end{equation}

and get the commutation rule $[Q(t),P(t)]=i\hbar$. The eigenfunction
of $P(t)$ can be calculated in two ways: (1) we can calculate the
eigenvector of $P(t)$ in the $Q_{0}$ representation using Eq.(\ref{Pdef})
and then use Green's function Eq.(\ref{greens_function}) to transform
it into the $Q(t)$ representation and (2) the commutation rule $[Q(t),P(t)]=i\hbar$
requires that $P(t)=-i\hbar\frac{\partial}{\partial Q}$, so the eigenfunction
of $P(t)$ with eigenvalue $P_{1}$ is $\exp[i\frac{P_{1}}{\hbar}Q]$.
By comparing these two solutions, the arbitrary phase $\phi(Q_{1},t)$
in Green's function is determined to be within a phase $\phi(t)$
, which is independent of $Q_{1}$. $\phi(t)$ is an arbitrary real
function of time, except that $\phi(0)$ = 0 so that it satisfies
the condition that at $t=0$ , the Green's function becomes $\delta(Q_{1}-Q_{0})$.
Thus, we obtain Green's function in the $S_{Q}$ space:

\begin{align}
 & G(Q_{1},Q_{0};t,0)=\left(\frac{M\omega e^{\frac{\eta}{2}t}}{2\pi i\hbar\sin(\omega t)}\right)^{\frac{1}{2}}\exp\left[i\frac{M}{2\hbar a_{2}}\left(a_{1}Q_{0}^{2}-2Q_{1}Q_{0}+e^{\eta t}\dot{a}_{2}Q_{1}^{2}\right)-\frac{i}{\hbar}\phi(t)\right]
\end{align}

It is then straightforward to derive the Hamiltonian $H_{Q}$ using
the following relation:

\[
H_{Q}=i\hbar\left(\frac{\partial}{\partial t}U_{Q}\right)U_{Q}^{-1}
\]

and remember that the matrix elements of $U_{Q}$ and $U_{Q}^{-1}$
are Green's function and its conjugate. The result is

\begin{equation}
H_{Q}=e^{-\eta t}\frac{P^{2}}{2M}+\frac{1}{2}Me^{\eta t}\omega_{0}^{2}Q^{2}+\dot{\phi}(t)\label{effective_Hamiltonian}
\end{equation}
Since $\phi$ is arbitrary except that $\phi((0)=0$, we can take
$\phi(t)=0$. Therefore, we have derived the well-known effective
Hamiltonian for the dissipative system. We emphasize that the expression
for $H_{Q}$ is derived here, whereas it is usually introduced by
heuristic arguments.

Next, we shall analyze the effect of the bath. Similar to Eq.(\ref{Qcd}),
we define the contribution of the bath oscillator $j$ to the Brownian
motion of the main oscillator as

\[
\text{\ensuremath{\xi_{j}(t)\equiv x_{j0}}}b_{j1}(t)+\text{\ensuremath{\dot{x}_{j0}}}b_{j2}(t)
\]

where $b_{j1}(t),b_{j1}(t),\dot{b}_{j1}(t),\dot{b}_{j1}(t)$ all equal
to zero at $t=0$ (see Appendix I), so $\xi_{j}(t=0)=0$ and $\dot{\xi}_{j}(t=0)=0$,
and hence in Eq.(\ref{Qxi_def-1}), $q(0)=q_{0},\dot{q}(0)=\dot{q}_{0}$.

Similar to Eq.(\ref{uQ}) we obtain the eigenfunctions $\theta_{\xi_{j1}}(x_{j0},t)$
for $\xi_{j}$ and $t>0$. Using Dirac's notation we have

\begin{align*}
 & Q|u_{Q_{1}},s\rangle=Q_{1}|u_{Q_{1}},s\rangle\\
 & \xi_{j}|\theta_{\xi_{j1}}\rangle=\xi_{j_{1}}|\theta_{\xi_{j1}}\rangle,
\end{align*}

where $s=\pm1$ is the eigenvalue of $\sigma_{z}$: 

\[
\sigma_{z}|s>=s|s>,
\]

and 

\[
|u_{Q},s\rangle\equiv u_{Q}(Q_{0},t)|s>.
\]

Thus $|u_{Q},s\rangle\prod_{j}\otimes|\theta_{\xi_{j}}\rangle$ is
an eigenvector of $q(t)$, with eigenvalue of $q=Q+sd+\sum_{j}\xi_{j}$.
Hence $Q=q-sd-\sum_{j}\xi_{j}$. In other words, the eigenvector of
$q(t)$ with eigenvalue $q$ is

\begin{equation}
|q,s,\{\xi_{j}\}\rangle=|u_{q-sd-\sum_{j}\xi_{j}},s\rangle\prod_{j}\otimes|\theta_{\xi_{j1}}\rangle\label{system_eigenstate}
\end{equation}

The set of Hermitian $q,s,\{\xi_{j}\}$ is a set of commuting observables
distinct from the set of $q,s,\{x_{j}$\}. The set of wave functions
$|q,s,\{\xi_{j}\}\rangle$ forms a Hilbert space's orthonormal basis.

We first study the evolution of the damped simple harmonic oscillator
wave function. Initially ($t=0$) it is $u_{Q_{1}}(Q_{0},t)$. Then,
at time $t$, according to the discussion following Eq.(\ref{uQ})
, $G(Q_{1},Q_{0};t,0)\equiv\left\langle Q_{1}\left|U_{Q}(t)\right|Q_{0}\right\rangle $,
and $u_{Q_{1}}(Q_{0},t)=\left\langle Q_{0}\left|U_{Q}^{-1}(t)\right|Q_{1}\right\rangle $,
it evolves into

\begin{align*}
 & \intop dQ_{0}G(Q_{2},Q_{0};t,0)u_{Q_{1}}(Q_{0},t)=\intop dQ_{0}\left\langle Q_{2}\left|U_{Q}(t)\right|Q_{0}\right\rangle \left\langle Q_{0}\left|U_{Q}^{-1}(t)\right|Q_{1}\right\rangle \\
 & =\left\langle Q_{2}\left|U_{Q}(t)U_{Q}^{-1}(t)\right|Q_{1}\right\rangle =\left\langle Q_{2}\mid Q_{1}\right\rangle =\delta(Q_{2}-Q_{1})
\end{align*}

Similarly, we can confirm that if the wave function of the bath oscillator
$j$ is $\theta_{\xi_{j1}}(x_{j0},t)$, at time $t>0$, it evolves
into $\delta(\xi_{j}-\xi_{j1})$ at time $t>0$.

We emphasize that if the initial state (at $t=0_{+}$) is $|q_{1},s_{1},\{\xi_{j1}\}\rangle$,
then the wave function will evolve at time $t$ into $\delta(q-q_{1})\delta_{s,s_{1}}\prod_{j}\delta(\xi_{j}-\xi_{j1})$.
At time $t=0_{+}$, because $Q(t),$$\sigma_{z}$ and $\{\xi_{j}(t)\}$
are linear combinations of operators $q_{0},\dot{q}_{0},\sigma_{z},\text{\ensuremath{x_{j0}}},\text{\ensuremath{\dot{x}_{j0}}}$,
they form a complete set of commuting observables of time $t=0_{+}$.
In other words, at time $t=0_{+}$, we consider the unitary transform
from $q_{0},\sigma_{z},\text{\ensuremath{x_{j0}},}$ to $Q(t),$$\sigma_{z}$
and $\{\xi_{j}(t)\}$ as a variable transform. The set of wave functions
$|q_{1},s_{1},\{\xi_{j1}\}\rangle$ forms an orthonormal basis in
the system's Hilbert space, which we call the \textquotedbl eigenstates
of the system''. As a result, if the system is initially (that is,
at $t=0_{+}$) in one of these states, $|q_{1},s_{1},\{\xi_{j1}\}\rangle$,
the final result at time $t$ is predetermined, i.e., the measurement
is deterministic, with a definite value of $q_{1},s_{1},\{\xi_{j1}\}$
.

\section*{2. The initial superposition of eigenstates determines the final
probability distribution}

If the wave function is initially $|\Psi_{0}\rangle=|\psi_{0},s\rangle\prod_{j}\otimes|\chi_{j0}\rangle$,
then to calculate the wave function at time $t$, we must expand $\Psi_{0}$
in terms of the eigenvectors $|q,s,\{\xi_{j}\}\rangle$ in Eq.(\ref{system_eigenstate}),
i.e., we must calculate

\begin{align}
 & \psi(Q,s,t)=<u_{Q},s,t|\psi_{0},s>=\int u_{Q}^{*}(q_{0},t)\psi_{0}(q_{0})dq_{0}<s|s>=\int u_{Q}^{*}(q_{0},t)\psi_{0}(q_{0})dq_{0}\label{psi_xi}\\
 & \chi_{j}(\xi_{j},t)=<\theta_{\xi_{j}},t|\chi_{j0}>=\int\theta_{\xi_{j}}^{*}(x_{j0},t)\chi_{j0}(x_{j0})dx_{j0}\nonumber 
\end{align}

Notice that even though we label these functions by the parameter
$t$, they are the wave functions at time $t=0$. We label them with
$t$ only to show that we define them as the initial state that will
evolve to specified $Q,\{\xi_{j}\}$ at time $t$. The probability
amplitude for the system in state $|q_{1},s_{1},\{\xi_{j1}\}\rangle$
at time $t=0_{+}$ is then

\begin{align}
 & \Psi(q_{1},s_{1},\{\xi_{j1}\},t)=<q_{1},s_{1},\{\xi_{j1}\}|\Psi_{0}>=<u_{q_{1}-s_{1}d-\sum_{j}\xi_{j1}},s_{1},t|\psi_{0}>|s_{1}>\prod_{j}<\theta_{\xi_{j1}},t|\chi_{j0}>\nonumber \\
 & \equiv\psi(q_{1}-s_{1}d-\sum_{j}\xi_{j1},s_{1},t)\prod_{j}\chi_{j}(\xi_{j1},t)\label{psi_projection1}
\end{align}

Because $|q_{1},s_{1},\{\xi_{j1}\}\rangle$ evolves into $\delta(q-q_{1})\delta_{s,s_{1}}\prod_{j}\delta(\xi_{j}-\xi_{j1})$
at time $t$, the wave function at time $t$ is

\begin{align}
 & \int dq_{1}\prod_{j}\int d\xi_{j1}\sum_{s_{1}}\delta(q-q_{1})\delta_{s,s_{1}}\prod_{j}\delta(\xi_{j}-\xi_{j1})\Psi(q_{1},s_{1},\{\xi_{j1}\},t)=\Psi(q,s,\{\xi_{j}\},t)\label{psi_projection2}
\end{align}

This equation seems to be redundant because it simply replaces $q_{1},s_{1},\{\xi_{j1}\}$
by $q,s,\{\xi_{j}\}$, so in the following, we shall only use Eq.(\ref{psi_projection1})
to get the wave function. However, the purpose of this redundant step
is to point out that the final wave function is obtained by projecting
the initial state onto the system eigenstates, which themselves are
taken as the initial states. We emphasize here again that this probability
amplitude is completely determined by the initial wave function $|\Psi_{0}\rangle$.
More examples will be provided to demonstrate the implications of
this point.

Notice that $\psi(Q,t)$ of Eq.(\ref{psi_xi}) is the wave function
in the Schrödinger representation with the effective Hamiltonian Eq.(\ref{effective_Hamiltonian}).
Thus, we have connected the effective Hamiltonian approach to the
dissipative system problem with the other approaches that take both
the system and the bath into account. We also notice that, despite
the fact that our $\Psi(q,s,\{\xi_{j}\},t)$ is in a different representation
than $\Psi(q,s,\{x_{j}\},t)$, the usual probability interpretation
remains valid: $\int\int...\int|$$\Psi(q,s,\{\xi_{j}\},t)|^{2}\prod_{j}d\xi_{j}$
is the probability density of finding the particle at $q$. Since
this solution is very simple, it provides a simple way to analyze
other complicated problems, e.g., studying the influence of Brownian
motion on interference, on which we shall not elaborate.

Under certain conditions, for example, at low temperatures and when
the system $q$ is in highly excited states, the range of $q$ is
large enough that we can approximately write $\Psi(q,s,\{\xi_{j}\},t)=\psi(q-sd,s,t)\prod_{j}\chi_{j}(\xi_{j},t)$
for all the values of $j$ that do not have a vanishingly small probability,
$q\gg|\sum_{j}\xi_{j}|$. That is, the wave function is factorized,
the dissipative system $q$ can be described by the wave function
$\psi(q-sd,s,t)$ only, and the Brownian motion can be ignored. As
a result, it is worthwhile to investigate the width of the argument
of the wave function due to Brownian motion, i.e., the mean value
of $\left(\sum_{j}\xi_{j}\right)^{2}$ at time $t$. It can be calculated
using the expression $\psi(Q,s,t)$ and $\chi_{j}(\xi_{j},t)$ in
Eq.(\ref{psi_xi}), as shown in the example below.

As one example of the superposition of the ``eigenstates of the system,''
we assume the absolute temperature is zero; the initial state of the
bath oscillators is in the ground state; the initial state of the
damped simple harmonic oscillator is a wave packet at the origin ($q=0$)
with width same as the width of ground state of the simple harmonic
oscillator $\sigma_{q}^{2}=\frac{\hbar}{2M\omega_{0}}$ and the spin
state is $|s>=|+1>$. The initial state of the damped harmonic oscillator
is

\[
{\displaystyle \psi_{0}(Q_{0})=\left(\frac{1}{2\pi\sigma_{q}^{2}}\right)^{\frac{1}{4}}\exp\left(-\frac{1}{4\sigma_{q}^{2}}\left(Q_{0}+d\right)^{2}\right)}
\]

where $Q_{0}=q_{0}-d$ and $d\equiv\frac{B}{\omega_{0}^{2}}$. Following
the step of Eq.(\ref{psi_projection1}) by Gaussian integration, we
find the wave function at time $t$ as a Gaussian,

\begin{align*}
 & \psi(Q,t)=\phi_{0}(Q,t)\exp\left(\frac{M\omega_{0}}{\hbar}\frac{-2Qd-a_{1}d^{2}}{2\left(a_{1}+i\omega_{0}a_{2}\right)}\right)\\
 & \phi_{0}(Q,t)=\left(\frac{M\omega_{0}}{\pi\hbar}\right)^{\frac{1}{4}}\left(\frac{1}{a_{1}(t)+i\omega_{0}a_{2}(t)}\right)^{\frac{1}{2}}\exp\left(-\frac{M\omega_{0}}{2\hbar}\frac{\omega_{0}a_{2}(t)-i\dot{a}_{2}(t)}{\omega_{0}a_{2}(t)-ia_{1}(t)}e^{\eta t}Q^{2}\right)
\end{align*}

where $\phi_{0}(Q,t)$ is the solution when the magnetic field is
turned off ($B=0$). The functions $a_{1}(t),a_{1}(t),\dot{a}_{1}(t),\dot{a}_{2}(t),$are
given in Appendix I. The initial function $\psi_{0}(Q_{0})=\phi_{0}(Q_{0}+d,t=0)$
is the function of the ground state of the oscillator with the centroid
displaced to $-d$ but $Q_{0}$ is the position relative to $d$,
so the centroid is at $q_{0}=0$.

The probability distribution of $Q$ ignoring the contribution from
the bath, is given by

\begin{align}
 & |\psi(Q,t)|^{2}=\left(\frac{1}{2\pi\sigma_{Q}^{2}}\right)^{\frac{1}{2}}\exp\left(-\frac{\left(Q+a_{1}d\right)^{2}}{2\sigma_{Q}^{2}}\right)\label{psiQ}\\
 & \sigma_{Q}^{2}\equiv\frac{\hbar}{2M\omega_{0}}\left(\omega_{0}^{2}a_{2}^{2}+a_{1}^{2}\right)=\sigma_{q}^{2}\left(1+\frac{\eta^{2}}{4\omega^{2}}+\frac{1}{2}\frac{\eta}{\omega}\sin(2\omega t)-\frac{\eta^{2}}{4\omega^{2}}\cos(2\omega t)\right)e^{-\eta t}\nonumber 
\end{align}

This is a wave packet that begins at $Q=-d$ (i.e., $q=0$), oscillates
around $Q=0$ (i.e. $q=d)$ , and has its amplitude dampened to zero
because $a_{1}(t)=e^{-\frac{\eta}{2}t}\left(\cos(\omega t)+\frac{\eta}{2\omega}\sin(\omega t)\right)$.
With damping time $2/\eta$, the width $\sigma_{Q}(t)$ also damps
to zero,so the wave function collapses to the point at $d$. If the
initial spin state is $s=-1$, it collapses to $-d$.

To include the Brownian motion from the bath, we have the bath oscillator
initial wave function in ground state

\[
{\displaystyle \chi_{j0}(x_{j0})=\left(\frac{m_{j}\omega_{j}}{\pi\hbar}\right)^{\frac{1}{4}}\exp\left(-\frac{m_{j}\omega_{j}}{2\hbar}x_{j0}^{2}\right)}
\]

The eigenfunction of $\xi_{j}|\theta_{\xi_{j1}}\rangle=\xi_{j_{1}}|\theta_{\xi_{j1}}\rangle$
gives $\theta_{\xi_{j1}}(x_{j0},t)=\left(\frac{m_{j}}{2\pi\hbar b_{j2}(t)}\right)^{\frac{1}{2}}\exp\left[-i\frac{m_{j}}{2\hbar b_{j2}(t)}\left(b_{j1}(t)x_{j0}^{2}-2\xi_{j1}x_{j0}\right)\right]$.
The derivation of $\theta_{\xi_{j1}}(x_{j0},t)$ is very similar to
the derivation of the eigenfunction $u_{Q}$ of $Q(t)$ in Appendix
II, (see Appendix I for the expression for $b_{j1}(t),b_{j2}(t)$
). The projection of the ground state onto the eigenvector in Eq.(\ref{psi_projection1})
is found by another Gaussian integral as

\begin{align}
 & \chi_{j}(\xi_{j},t)=<\theta_{\xi_{j}},t|\chi_{j0}>=\left(\frac{m_{j}\omega_{j}}{\pi\hbar}\right)^{\frac{1}{4}}\left(\frac{1}{-ib_{j1}+\omega_{j}b_{j2}(t)}\right)^{\frac{1}{2}}\exp\left(-\frac{m_{j}}{2\hbar b_{j2}}\frac{\xi_{j1}^{2}}{\left(-ib_{j1}+\omega_{j}b_{j2}(t)\right)}\right)\label{xi_j}
\end{align}

As a result, the contribution to the Brownian motion width derived
from $\chi_{j}(\xi_{j},t)$ is $<\xi_{j}^{2}>=\sigma_{\xi_{j}}^{2}=\frac{1}{2}\left(b_{j1}^{2}+\omega_{j}^{2}b_{j2}^{2}\right)\frac{\hbar}{m_{j}\omega_{j}}$.

According to the discussion that follows Eq.(\ref{psi_projection1},\ref{psi_projection2}),
the width of the wave function

\[
\Psi(q_{1},s_{1},\{\xi_{j1}\},t)=\psi(q-sd-\sum_{j}\xi_{j},s,t)|s>\prod_{j}\chi_{j}(\xi_{j},t)
\]

can be calculated from the widths of $\psi(Q,t)$ in Eq.(\ref{psiQ})
and $\chi_{j}(\xi_{j},t)$ in Eq.(\ref{xi_j}) as $\sigma_{Q}^{2}+<\left(\sum_{j}\xi_{j}\right)^{2}>$
(see Appendix III for the derivation). Long after the damping time
$2/\eta$ , the width $\sigma_{Q}^{2}\rightarrow0$ , and thus the
Brownian width is given by

\[
<\left(\sum_{j}\xi_{j}(t)\right)^{2}>=\int\int...\int\left(\sum_{j}\xi_{j}(t)\right)^{2}\prod_{j}|\chi_{j}(\xi_{j},t)|^{2}d\xi_{j}=\sum_{j}<\xi_{j}^{2}>=\sum_{j}\sigma_{\xi_{j}}^{2}
\]

From the bath's spectral density Eq.(\ref{rho_def}), and the expressions
for $b_{j1},b_{j2},\omega,\omega_{0},\eta$ in Appendix I, and Eq.(\ref{qxsol-1}),
we find

\begin{align}
 & \sigma_{\xi}^{2}\equiv\sum_{j}\sigma_{\xi_{j}}^{2}=\sum_{j}\frac{\hbar}{2m_{j}\omega_{j}}\left(b_{j1}^{2}+\omega_{j}^{2}b_{j2}^{2}\right)=\int d\omega_{j}\rho(\omega_{j})\frac{\hbar}{2m_{j}\omega_{j}}\left(b_{j1}^{2}+\omega_{j}^{2}b_{j2}^{2}\right)\label{Brownian_width}\\
 & =\frac{\eta\hbar}{\pi M}\int d\omega_{j}\omega_{j}\frac{M^{2}}{c_{j}^{2}}\left(b_{j1}^{2}+\omega_{j}^{2}b_{j2}^{2}\right)\nonumber 
\end{align}

As $t\rightarrow\infty$, this approaches

\begin{align*}
 & \sigma_{\xi}^{2}=\sum_{j}\sigma_{\xi_{j}}^{2}=\frac{\eta\hbar}{2\pi M}\int_{0}^{\infty}d\omega_{j}^{2}\frac{1}{\text{(\ensuremath{\omega_{0}^{2}}-\ensuremath{\omega_{j}^{2})^{2}}+\ensuremath{\eta^{2}\omega_{j}^{2}}}}=\frac{\hbar}{2\pi\omega M}\left(\frac{\pi}{2}+\arctan(\frac{\omega^{2}-\frac{\eta^{2}}{4}}{\eta\omega})\right)
\end{align*}

If $\eta\ll$$\omega$, i.e., if the damping time is much longer than
the main oscillator period.

\begin{align*}
 & <\left(\sum_{j}\xi_{j}(t)\right)^{2}>\approx\frac{\hbar}{2\pi\omega M}\left(\frac{\pi}{2}+\arctan(\frac{\omega}{\eta})\right)\approx\frac{\hbar}{2\omega M}
\end{align*}

This Brownian width approximately equals the ground state of the simple
harmonic oscillator.

As a result, depending on whether $s=1$ or $-1$, the wave function
collapses to one of the two split potential wells at $\pm d$ with
a spread $\sigma_{\xi}=\sqrt{\frac{\hbar}{2\omega M}}$. As long as
$d\gg\sqrt{\frac{\hbar}{2\omega M}}$ , the spin measurement has a
definite answer. From the discussion following Eq.(\ref{psi_projection1},\ref{psi_projection2}),
this result is completely predetermined by the initial spin state.
\begin{figure}[t]
\includegraphics[viewport=0bp 0bp 360bp 216bp,clip,scale=0.6]{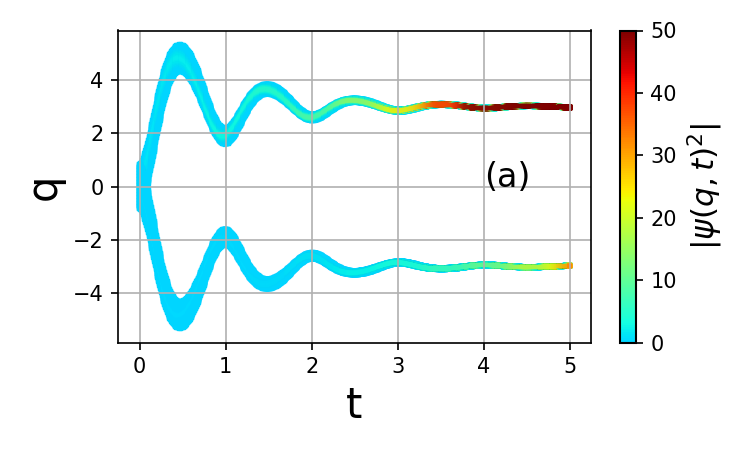}\vspace{-0.0em}\includegraphics[scale=0.6]{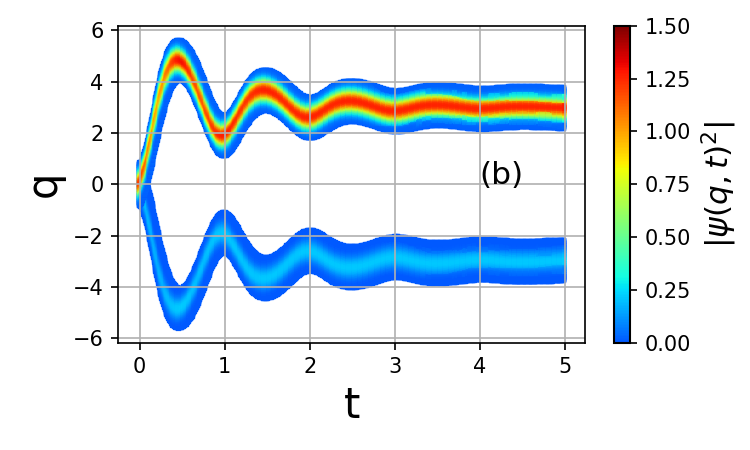}\vspace{-1.0em}

\caption{(a) Wave function probability density without Brownian motion (b)
with Brownian motion. The initial spin state is specified by a vector
$\vec{a}$ on the Block sphere with $\theta=\pi/4$ for this case.
The probability for the spin up and down is both non-zero. The color
scale is the probability density. The two wave packets of the spin
up and down are well separated near the beginning. When Brownian motion
is ignored, the wave function profile width approaches zero at $t\protect\geq5$.
The wave function collapses to either at $q=d=3$, or $q=-d=-3$ for
$t>5$, with the probability of being up and down as $\cos^{2}(\theta/2)=0.8535$
, and $\sin^{2}(\theta/2)=0.1464$.}
\vspace{-1.0em}
\end{figure}

\section*{3. An example of the case when the Bloch vector of the pure state
spinor is not aligned with the $z$ axis}

When the spin is in a pure state, the initial state can always be
specified by a vector $\vec{a}=(\sin\theta\cos\phi,\sin\theta\sin\phi,\cos\theta)$
on the Bloch sphere\citep{bloch} as

\begin{align*}
 & |\vec{a}>=\cos(\theta/2)\left|+\right\rangle +e^{i\phi}\sin(\theta/2)\left|-\right\rangle 
\end{align*}

Then the initial state $\Psi_{0}=|\psi_{0}>|\vec{a}>\prod_{j}|\chi_{j0}>$.

$\vec{a}$ is not parallel to the $z$ axis unless $\theta=0$, or
$\pi$. Its projection onto the initial system eigenstate $|q_{1},s_{1},\{\xi_{j1}\}>$
in Eq.(\ref{system_eigenstate}) , given by Eq.(\ref{psi_projection1},\ref{xi_j})
is

\begin{align*}
 & \Psi(q_{1},s_{1},\{\xi_{j1}\},t)=<q_{1},s_{1},\{\xi_{j1}\}|\Psi_{0}>=<u_{q_{1}-s_{1}d-\sum_{j}\xi_{j1}},s_{1},t|\psi_{0}><s_{1}|\vec{a}>\prod_{j}<\theta_{\xi_{j1}},t|\chi_{j0}>\\
 & =\psi(q_{1}-s_{1}d-\sum_{j}\xi_{j1},s_{1},t)<s_{1}|\vec{a}>\prod_{j}\chi_{j}(\xi_{j1},t)
\end{align*}

\begin{figure}[b]
\includegraphics[scale=0.4]{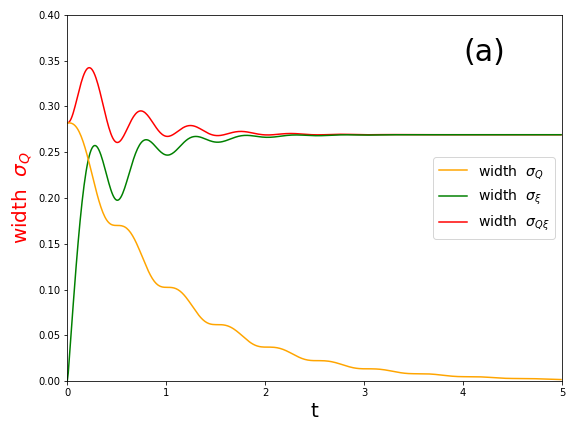}\vspace{-1.0em}\includegraphics[scale=0.4]{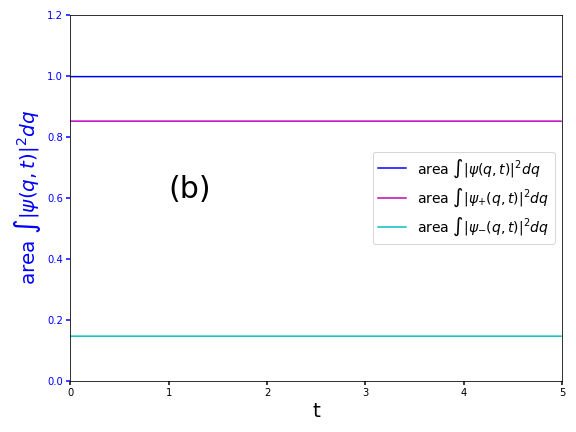}\vspace{-1.0em}

\caption{\textcolor{black}{(a) the width $\sigma_{Q\xi}(t)$, $,\sigma_{\xi}(t),$
and $\sigma_{Q}(t)$; (b)Probability in the wave packet with spin
up (magenta), down(cyan), and total (blue);}}
\vspace{-1.0em}
\end{figure}
\begin{figure}[t]
\includegraphics[scale=0.4]{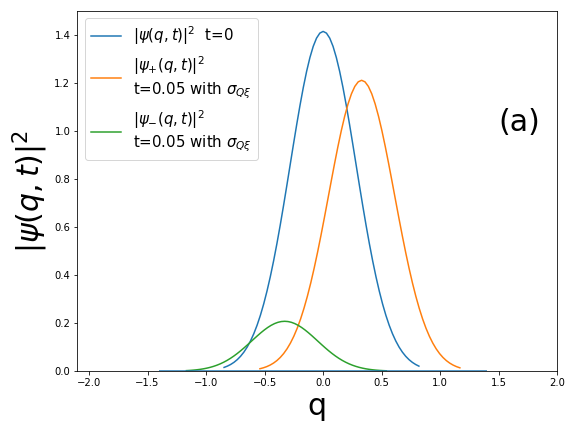}\vspace{-1.0em}\includegraphics[scale=0.4]{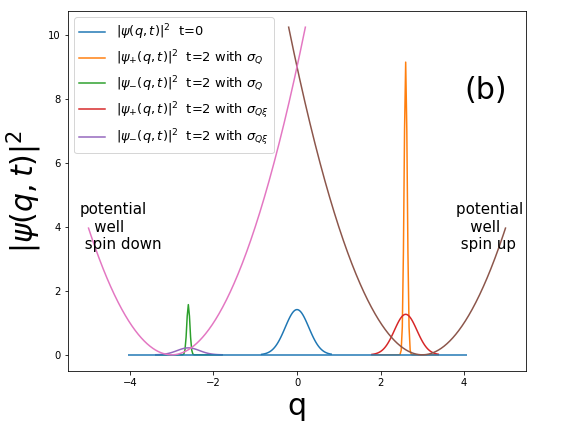}\vspace{-0.0em}

\caption{\textcolor{black}{(a) the wave function density profile of the wave
packets at $t=0$ and very close to the beginning at $t=0.05$. The
two wave packets are already starting to separate. (b) Wave function
density profile at $t=0$ and $t=2$ compares with and without Brownian
motion included. We also indicate the positions of the two potential
wells corresponding to spin up or down, with an arbitrary scale. The
two wave packets are close but do not reach the bottom of the potential
wells yet at $t=2$.}}
\vspace{-1.0em}
\end{figure}
According to the discussion following Eq.(\ref{psi_projection1},\ref{psi_projection2}),
at time $t$, $\Psi_{0}$ evolves into the wave function

\begin{align}
 & \Psi_{+}(q,t)\equiv\Psi(q,1,\{\xi_{j}\},t)=\psi(q-d-\sum_{j}\xi_{j},1,t)\cos(\theta/2)\prod_{j}\chi_{j}(\xi_{j},t)\label{psiplus}\\
 & \Psi_{-}(q,t)\equiv\Psi(q,-1,\{\xi_{j}\},t)=\psi(q+d-\sum_{j}\xi_{j},-1,t)e^{i\phi}\sin(\theta/2)\prod_{j}\chi_{j}(\xi_{j},t)\nonumber 
\end{align}

The wave packet split into two, with $s=1$ collapsing to $q=d$ with
a probability of $\cos^{2}(\theta/2)$, and $s=-1$ collapsing to
$q=-d$ with a probability of $\sin^{2}(\theta/2)$, both with a width
$\sigma_{\xi}=\sqrt{\frac{\hbar}{2\omega M}}$, given by Eq.(\ref{Brownian_width})

At temperature $T$ the contribution from the Brownian width of the
bath to the width of $q$ is calculated by the bath density matrix
as

\[
<\left(\sum_{j}\xi_{j}(t)\right)^{2}>=\sum_{j}\frac{\hbar}{2m_{j}\omega_{j}}\left(b_{j1}^{2}+\omega_{j}^{2}b_{j2}^{2}\right)\coth(\frac{\hbar\omega_{j}}{2kT})
\]

This approaches the width in Eq.(\ref{Brownian_width}) as $T$ approaches
zero. The width increases with temperature.

When we apply Eq.(\ref{wave_density}) and Eq.(\ref{psiQ}) for $|\psi(Q,t)|^{2}$
given in Appendix III, we have

\begin{align*}
 & |\Psi_{+}(q,t)|^{2}=\cos^{2}(\theta/2)\frac{1}{\sqrt{2\pi}}\frac{1}{\sigma_{Q\xi}}\exp\left(-\frac{\left(q-d+a_{1}d\right)^{2}}{2\sigma_{Q\xi}^{2}}\right)\\
 & |\Psi_{-}(q,t)|^{2}=\sin^{2}(\theta/2)\frac{1}{\sqrt{2\pi}}\frac{1}{\sigma_{Q\xi}}\exp\left(-\frac{\left(q+d-a_{1}d\right)^{2}}{2\sigma_{Q\xi}^{2}}\right)
\end{align*}

where the width of the wave function density is $\sigma_{Q\xi}^{2}=\sigma_{Q}^{2}+\sigma_{\xi}^{2}$.

For this abstract model, for a case with $\hbar=1,M=1$, and the main
oscillator period $T_{0}=1,\omega_{0}=2\pi/T_{0}$, $\eta=2$, we
choose $d=3$, at temperature $T=0$. The angle between the spin direction
and the $z$ axis is chosen to be $\theta=\pi/4$. We plot the $|\Psi_{+}(q,t)|^{2},|\Psi_{-}(q,t)|^{2}$
versus $t$ in Fig.1a when the Brownian motion is ignored, i.e. with
width $\sigma_{Q}$, and in Fig.1b with Brownian motion included,
i.e., with width $\sigma_{Q\xi}$. The separation of the two wave
packets is visible from the very beginning near $t=0$.

The width in Eq.(\ref{psiQ}) $\sigma_{Q}^{2}(t)\rightarrow0$ as
$t\rightarrow\infty$. As a result, the wave packet is already approaching
$\delta$ function at $t=5$ in Fig.1a.

The width of $|\psi(Q,t)|^{2}$, i.e., $\sigma_{Q}(t)$, Brownian
width $\sigma_{\xi}(t)$ and total width $\sigma_{Q\xi}(t)$, as function
of $t$ are also shown in Fig.2a.

The probability of falling into an upper and lower potential well
is shown in Fig.2b, i.e. they are constants starting from $t=0_{+}$:
for $\theta=\pi/4$, they are $\cos^{2}(\theta/2)=0.8535$ (magenta);
$\sin^{2}(\theta/2)=0.1464$ (cyan). That is, they are predetermined.
Their sum is $1$(blue). 

The profile of wave packets with spin up and down begins to separate
at $t=0.05$, as shown in Fig.3a. The wave packets at $t=2$ are shown
in Fig.3b, where we compare the profiles with and without Brownian
motion when they are well separated.

The conclusion is: the probability distribution $\cos^{2}(\theta/2)$,
$\sin^{2}(\theta/2)$ is predetermined by the initial state $|\vec{a}>$.

\section*{4. Implications for applying the result to the Bell theorem}

We can now apply the results of Sections 1,2,3 to the Bell theorem.
The main results of these sections are that the final result of the
measurement is determined by the wave function of the initial state,
as stressed at the end of Section 1, and clearly shown in Fig.2b.
Since this is only for a specific model, the interpretation and the
implication here are only suggestive.

\subsubsection*{4.1. We first repeat the statement of the Bell theorem to clarify
the notations used here.}

In the Bell theorem experiment \citep{bell,bell_theorem}, we assume
that the two entangled particles leave the origin at point O and move
in opposite directions, so that particle 1 reaches position A and
measures $\sigma_{1}\centerdot\vec{a}$ and particle 2 reaches position
B and measures $\sigma_{2}\centerdot\vec{b}$ (\textcolor{black}{$\vec{a}$
,$\vec{b}$ }are some unit vectors).

The paper assumed a hidden variable $\lambda$ such that the result
of measuring $\sigma_{1}\centerdot\vec{a}$ is determined by $\vec{a}$
and $\lambda$ as $A(\vec{a},\lambda)$, and the result of measuring
$\sigma_{2}\centerdot\vec{b}$ is determined by $\vec{b}$ and $\lambda$
as $B(\vec{b},\lambda)$ in the same instance, and

\begin{equation}
A(\vec{a},\lambda)=\pm1;B(\vec{b},\lambda)=\pm1.\label{AB1}
\end{equation}
According to Bell, ``The vital assumption is that the result B for
particle 2 does not depend on the setting $\vec{a}$ of the magnet
for particle 1, nor A on $\vec{b}$''\citep{bell}. Bell assumed
that if $\rho(\lambda)$ is the probability distribution of $\lambda$,
then the expectation value of the product of the two components $\sigma_{1}\centerdot\vec{a}$
and $\sigma_{2}\centerdot\vec{b}$ is

\begin{equation}
P(\vec{a},\vec{b})=\intop d\lambda\rho(\lambda)A(\vec{a},\lambda)B(\vec{b},\lambda)\label{classical_probability}
\end{equation}

Bell supposes that the experimenter has a choice of settings for the
second detector: it can be set either to $\vec{b}$ or to $\vec{c}$
. Bell proves that if Eq.(\ref{AB1}) and Eq.(\ref{classical_probability})
are correct, the difference in correlation between these two choices
of detector settings must satisfy the inequality

\[
|P(\vec{a},\vec{b})-P(\vec{a},\vec{c})|\leq1+P(\vec{b},\vec{c})
\]

Then, Bell shows it is easy to find situations where quantum mechanics
violates the inequality, i.e., the ``Bell inequality'', thus Eq.(\ref{AB1})
and Eq.(\ref{classical_probability}) derived from the ``local hidden
variable theory,'' is incompatible with quantum mechanics.

\subsubsection*{4.2. The analysis of two separate measurements at remote positions
A and B on two particles is equivalent to the analysis of two measurements
on one particle at the origin}

When we apply the result of Sections 1, 2, and 3, because the result
is predetermined by the initial condition, it is clear that the measurement
at any later time would not affect the initial probability amplitude.
And, any conclusion from the result does not provide any information
about whether the measurement at $A$ influences $B$. Consequently,
the experiment does not carry any information about ``action at a
distance,'' or ``spooky action,'' or ``nonlocal'' property.

Therefore, we only need to compare the probability amplitude at time
$t=0$, i.e., we compare the measurement of $\sigma_{1}\centerdot\vec{a}$
and $\sigma_{2}\centerdot\vec{b}$ at time $t=0$. When we compare
these two measurements with the experiment in the Bell theorem, we
can compare them at $t=0$, which means we compare them at the origin
$O$.

Furthermore, because the two entangled particles are completely correlated
such that $\sigma_{1}\centerdot\vec{a}=-\sigma_{2}\centerdot\vec{a}$,
we only need to compare $-\sigma_{2}\centerdot\vec{a}$ with $\sigma_{2}\centerdot\vec{b}$
at time $t=0$ and investigate the condition for both to have definite
values. In the following, when applying the results of Section 3,
we can consider $\vec{z}$ as $\vec{b}$, and take $\vec{a}$ the
same as the $\vec{a}$ in Section 3.

\subsubsection*{4.3 The statement ``There are no hidden variables for non-commuting
observables to have definite values at the same time'' is the consequence
of the uncertainty principle. We discuss its implications for the
Bell theorem and ``Schr\textcyr{\"\cyro}dinger's cat problem''.}

Because we start with $\sigma_{1}\centerdot\vec{a}=\pm1$, and measure
$\sigma_{2}\centerdot\vec{b}$, the complete correlation between particles
1 and 2 requires particle 2's initial state to be either $\sigma_{2}\centerdot\vec{a}=1$,
or $\sigma_{2}\centerdot\vec{a}=-1$. So as long as $\sin\theta\neq0$,
particle 2's initial state is not an eigenstate of $\sigma_{2}\centerdot\vec{b}$.
Comparing with the Bell theorem, this is equivalent to measuring $\sigma_{2}\centerdot\vec{a}$
first and then immediately measuring $\sigma_{2}\centerdot\vec{b}$.

The initial state for the measurement of $\sigma_{2}\centerdot\vec{b}$
is then an eigenstate of $\sigma_{2}\centerdot\vec{a}$, not an eigenstate
of $\sigma_{2}\centerdot\vec{b}$, even though it is a pure state,
i.e., it is the result of the measurement of $\sigma_{2}\centerdot\vec{a}$.
As a result, as long as $\sin\theta\neq0$, the measurement of $\sigma_{2}\centerdot\vec{b}$
must have two possible outcomes. This is not determined by any external
observer, but is because the initial state is not a definite state
for the measurement of $\sigma_{2}\centerdot\vec{b}$, even though
it is a definite state for $\sigma_{2}\centerdot\vec{a}$.

It is also clear that if $\sin\theta\neq0$, there are no hidden variables
that can give both $\sigma_{2}\centerdot\vec{a}$ and $\sigma_{2}\centerdot\vec{b}$
definite values. This is the Uncertainty Principle. Further, the probability
of $\sigma_{2}\centerdot\vec{b}=1$ is $\cos^{2}(\theta/2)$, and
the probability of $\sigma_{2}\centerdot\vec{b}=-1$ is $\sin^{2}(\theta/2)$
according to the Born rule.

In the present model case, it is very natural to understand why if
$\sin\theta\neq0$, there are two outcomes from $\sigma_{2}\centerdot\vec{b}$.
Since the initial state is a pure state with $\vec{a}$ on the Bloch
sphere, it is a superposition of two components. The one with $\sigma_{2}\centerdot\vec{b}=1$
will be deflected by the magnetic field to direction $\vec{b}$, the
other with $\sigma_{2}\centerdot\vec{b}=-1$ will be deflected to
direction $-\vec{b}$. This probability distribution between the two
outcomes, on the other hand, exists from the start as a physical reality
embodied later by magnetic deflection.

As a result, the measurement of $\sigma_{2}\centerdot\vec{a}$ is
insufficient to prepare a definite state for the second measurement;
we must still specify that the measurement setup in the environment
after the start is still a measurement of $\sigma_{2}\centerdot\vec{a}$,
or in other words, a measurement of $\sigma_{2}\centerdot\vec{b}$
but with $\vec{b}=\pm\vec{a}$.

The implication of this is that even though the initial wave function
is the eigenvector of a complete set of commuting observables and
has definite values, this alone cannot guarantee a definite status;
because, in addition to this, we must first make sure the measurement
device is arranged to measure this same set of observables. From this
point of view, the wave function is not a complete description of
the system; a description of the measurement environment must be included
in order to have a complete description. One of the foundations of
quantum mechanics, the Uncertainty Principle, is based on the fact
that no physical system can be built to let two non-commuting observables
have definite values; consequently, no one can prepare such an initial
state in any experiment.

\textcolor{black}{If we consider the case in Section 3 with $\vec{a}=\vec{x}=(1,0,0)$,
then $\theta=\frac{\pi}{2}$,$\phi=0$,
\begin{align*}
 & |\vec{a}>=\cos(\theta/2)\left|+\right\rangle +e^{i\phi}\sin(\theta/2)\left|-\right\rangle =\frac{1}{\sqrt{2}}\left(\left|+\right\rangle +\left|-\right\rangle \right)
\end{align*}
}

\textcolor{black}{we may consider this as a model to compare with
an extremely simplified ``Schr\textcyr{\"\cyro}dinger's cat problem''.
It is not an eigenstate of $\sigma_{z}$ which is the measurement
we are taking. According to our discussion above, the initial state
is an eigenstate of the measurement of $\sigma\centerdot\vec{a}$.
It is not a definite state for the measurement of $\sigma_{z}$ from
the beginning, because it has a probability of 50\% for spin up or
down. So the initial measurement of spin $\sigma\centerdot\vec{a}$
produced an ensemble of initial states for the following measurement
of $\sigma_{z}$. We emphasize that this probability distribution
is not generated at the end by someone observing it when the measurement
of $\sigma_{z}$ is completed, but is already determined right after
the measurement of $\sigma\centerdot\vec{a}$. The role of a final
measurement of $\sigma_{z}$ is to sort out the distribution, sample
$\left|+\right\rangle $ or $\left|-\right\rangle $, and put them
into the centers of the two split potential wells.}

\textcolor{black}{Therefore, there is no need to resort to any ``external
observer''.}

\textcolor{black}{No matter whether there is any one to observe the
result, the probability distribution is certain from the very beginning.
$|\vec{a}>$ is called a ``pure state'' only in the sense that it
will give a definite result if one measures $\sigma\centerdot\vec{a}$.}\textcolor{red}{{}
}\textcolor{black}{The main point is no one can prepare a definite
state for $\sigma_{z}$ by measuring $\sigma\centerdot\vec{a}$.}

\subsubsection*{4.4. At this point, one may ask why the Bell theorem sometimes seems
to generate a sense of non-locality in quantum mechanics.}

As we discussed above, the violation of Bell inequality by quantum
mechanics is because the Uncertainty Principle contradicts the hidden-variable
theory. It was unnecessary to replace the ``hidden variable theory''
with the ``local hidden variable theory''. There is no need to resort
to the word ``local'' in the derivation of Bell inequality.

According to reference \citep{bell_theorem}, ``In the words of physicist
John Stewart Bell, for whom this family of results is named, 'If {[}a
hidden-variable theory{]} is local, it will not agree with quantum
mechanics, and if it agrees with quantum mechanics, it will not be
local' ''.

This conclusion seems to generate a sense of some non-locality in
quantum mechanics when the experiment indeed agrees with quantum mechanics.
However, with a careful examination of the derivation in 4.1, \textcolor{black}{as
long as we assume there is a hidden variable theory, i.e., Eq.(\ref{AB1})
and Eq.(\ref{classical_probability}), we can follow Bell's derivation
and reach the same conclusion as the Bell theorem without resorting
to locality or non-locality.}

\textcolor{black}{The the basic assumption by Bell for his derivation
is ``the result B for particle 2 does not depend on the setting $\vec{a}$
of the magnet for particle 1, nor A on $\vec{b}$''\citep{bell},
as we refered to in 4.1. We emphasize here that communication between
A and B (or not) is irrelevant. }

\textcolor{black}{Even if A does communicate with B such that B receives
information in advance that the setting chosen by A is $\vec{a}$,
as long as B does not choose the setting $\vec{b}$ to be parallel
to $\vec{a}$, Bell's assumption will still lead to Eq.(\ref{AB1})
and Eq.(\ref{classical_probability}), which then leads to Bell inequality
without resorting to locality or non-locality, i.e., without resorting
to whether there is any communication between A and B.}

\textcolor{black}{So we may rephrase the above statement (1) as a
new statement (2): }``If a hidden-variable theory can give definite
values to non-commuting observables, the result will not agree with
quantum mechanics, and if it agrees with quantum mechanics, then no
hidden-variable theory can give definite values to non-commuting observables.''

\textcolor{black}{As long as the Uncertainty Principle is correct,
it is natural that quantum mechanics will violate Bell inequality,
and also it rules out a ``hidden variable theory''. Both the Bell
theory and our discussion has nothing to do with whether there is
non-locality, ``action-at-a distance'', or not.}

\textcolor{black}{The theory we used is non-relativistic, which is
justified because there is nothing close to light speed involved.}

The comparison of \textcolor{black}{statement (1) and statement (2)
}and the discussion above demonstrates the experiment's agreement
with quantum mechanics does not provide information related to non-locality.
The inclusion of the word \textquotedbl local\textquotedbl{} in Bell's
statement here shifts the emphasis from \textquotedbl hidden variables
theory\textquotedbl{} to \textquotedbl local hidden variable theory\textquotedbl .
This contradicts quantum mechanics, giving the impression that the
confirmation of quantum mechanics prediction leads to some non-locality.
Hence, the discussion above suggests this statement may be misleading
if one does not carefully examine the basic ground underlying the
derivation of the Bell theorem.

We conclude that as long as the Uncertainty Principle is established,
violation of Bell inequality in the experiment entirely rules out
the existence of any hidden variable theory, irrelevant of any locality
or non-locality.

\section*{5. Summary: The implications for the understanding of the statistical
interpretation of quantum mechanics}

The preceding discussion in Section 4 appears to be simply repeating
the fundamentals of quantum mechanics; thus, as quantum mechanics
predicts, it leads to the violation of Bell Inequality. Indeed, we
have demonstrated that in quantum mechanics, when the Born Rule is
applied to the initial state, the result is a self-consistent theory
that includes using the Schr\textcyr{\"\cyro}dinger equation to describe
the quantum measurement process itself, without resorting to non-locality,
hidden variable, or any external observer; without resorting to whether
the measurement at A immediately influences the measurement at B or
the distance between A and B. Thus, it removes many mysteries surrounding
the statistical interpretation of quantum mechanics.

\textcolor{black}{We emphasize that because the discussion is only
for a specific model, the implications of the discussion in Section
4 are only suggestive and summarized as:}
\begin{enumerate}
\item \textcolor{black}{The derivation of the wave function collapse in
the measurement process in Sections 1, 2, and 3 shows the probability
distribution is determined from the very beginning (i.e., $t=0$),
not at the end.}
\item \textcolor{black}{As a result, when Bell inequality is violated in
an experimental test of the Bell theorem, there is no paradox arising
from the concepts of ``non-locality'' or ``action at a distance''.}
\item \textcolor{black}{A linear superposition of the eigenstates of a complete
set of commuting observables is an ensemble of systems with a probability
distribution determined by the Born rule. A wave function will give
a definite measurement result only if it is the eigenstate of the
observables being measured. This crucial point removes many mysteries
surrounding the statistical interpretation of quantum mechanics.}
\item This suggests that as long as the Born rule is applied to the interpretation
of the initial state, the Born rule at the end of the measurement
can be derived from the Schr\textcyr{\"\cyro}dinger equation.
\end{enumerate}
It is important to note, in the model case we have discussed here,
1) the correlation between the two entangled particles is assumed
to be completely preserved, and 2) the fact that the discussion about
the violation of Bell inequality without resorting to the distance
between A and B does not imply that the distance between A and B is
irrelevant to the experiment on the entangled particles. On the contrary,
the longer distance between A and B means the experiment design needs
to preserve the correlation between the two particles over a longer
distance.

We thank Prof. Y. Hao, Dr. T. Shaftan, Dr. V.Smaluk and Mrs. K. Stolle
for discussion and suggestions on the manuscript.
\begin{acknowledgments}
We thank Prof. Y. Hao, Dr. T. Shaftan, Dr. V.Smaluk and Mrs. K. Stolle
for discussion and suggestions on the manuscript.
\end{acknowledgments}

\section*{Appendix I}

The following expressions are used in the derivation in Sections 1,
2, and 3. In Eq.(\ref{qxsol-1}). with $a_{1}(t)=\frac{\mu e^{-\nu t}-\nu e^{-\mu t}}{\mu-\nu}$
, $a_{2}(t)=-\frac{e^{-\mu t}-e^{-\nu t}}{\mu-\nu}$,$\mu\equiv\frac{\eta}{2}+\omega i$,$\nu\equiv\frac{\eta}{2}-\omega i$
and here $\omega\equiv\left(\omega_{0}^{2}-\frac{\eta^{2}}{4}\right)^{1/2}$,
we have

\begin{align*}
 & a_{1}(t)=e^{-\frac{\eta}{2}t}\left(\cos(\omega t)+\frac{\eta}{2\omega}\sin(\omega t)\right)\\
 & a_{2}(t)=\frac{1}{\omega}e^{-\frac{\eta}{2}t}\sin(\omega t)\\
 & \dot{a}_{1}(t)=-\omega_{0}^{2}a_{2}(t)\\
 & \dot{a}_{2}(t)=e^{-\frac{\eta}{2}t}\cos(\omega t)-\frac{\eta}{2\omega}e^{-\frac{\eta}{2}t}\sin(\omega t)\\
 & a_{1}(t)\dot{a}_{2}(t)-\dot{a}_{1}(t)a_{2}(t)=e^{-\eta t}
\end{align*}

where we can confirm that $a_{1}(0)=1,a_{2}(0)=0,\dot{a}_{1}(0)=0,\dot{a}_{2}(0)=1$.

Then, use Eq.(\ref{Qcd}) and the expressions of $a_{1}(t),a_{12}(t),\dot{a}_{1}(t),\dot{a}_{2}(t)$,
we find the commutiator

\[
[Q,\dot{Q}]=[a_{1}(t)Q_{0}+a_{2}(t)\dot{Q}_{0},\dot{a}_{1}(t)Q_{0}+\dot{a_{2}}(t)\dot{Q}_{0}]=e^{-\eta t}\frac{i\hbar}{M}
\]

Also in the solution of the damped oscillator equation Eq.(\ref{dampin_eq}),
i.e., Eq.(\ref{qxsol-1}), the coefficients of the $j$'th bath oscillator
contribution to $q$

\begin{align*}
 & b_{j1}(t)=-\frac{c_{j}}{M}\text{InverseLaplace}(\frac{s}{\left(s^{2}+\omega_{j}^{2}\right)(s+\mu)(s+\nu)})\\
 & =-\frac{c_{j}}{M}\left[\frac{\omega_{j}(\mu+\nu)\sin(t\omega_{j})}{\left(\mu^{2}+\omega_{j}^{2}\right)\left(\nu^{2}+\omega_{j}^{2}\right)}+\frac{\left(\mu\nu-\omega_{j}^{2}\right)\cos(t\omega_{j})}{\left(\mu^{2}+\omega_{j}^{2}\right)\left(\nu^{2}+\omega_{j}^{2}\right)}+\frac{\mu e^{-\mu t}}{(\mu-\nu)\left(\mu^{2}+\omega_{j}^{2}\right)}-\frac{\nu e^{-\nu t}}{(\mu-\nu)\left(\nu^{2}+\omega_{j}^{2}\right)}\right]\\
 & b_{j2}(t)=-\frac{c_{j}}{M}\text{ InverseLaplace(\ensuremath{\frac{1}{\left(s^{2}+\omega_{j}^{2}\right)(s+\mu)(s+\nu)}})}\\
 & =-\frac{c_{j}}{M}\left[\frac{\left(\mu\nu-\omega_{j}^{2}\right)\sin(t\omega)}{\omega_{j}\left(\mu^{2}+\omega_{j}^{2}\right)\left(\nu^{2}+\omega_{j}^{2}\right)}-\frac{(\mu+\nu)\cos(t\omega_{j})}{\left(\mu^{2}+\omega_{j}^{2}\right)\left(\nu^{2}+\omega_{j}^{2}\right)}-\frac{e^{\mu(-t)}}{(\mu-\nu)\left(\mu^{2}+\omega_{j}^{2}\right)}+\frac{e^{\nu(-t)}}{(\mu-\nu)\left(\nu^{2}+\omega_{j}^{2}\right)}\right]\\
 & b_{j1}(t{\color{black}\rightarrow}\infty)=-\frac{c_{j}}{M}\left[\frac{\omega_{j}(\mu+\nu)\sin\left(t\omega_{j}\right)}{\left(\omega_{j}^{2}+\mu^{2}\right)\left(\omega_{j}^{2}+\nu^{2}\right)}+\frac{\left(\mu\nu-\omega_{j}^{2}\right)\cos\left(t\omega_{j}\right)}{\left(\omega_{j}^{2}+\mu^{2}\right)\left(\omega_{j}^{2}+\nu^{2}\right)}\right]\\
 & \omega_{j}b_{j2}(t{\color{black}\rightarrow}\infty)=-\frac{c_{j}}{M}\left[\frac{\left(\mu\nu-\omega_{j}^{2}\right)\sin\left(t\omega_{j}\right)}{\left(\omega_{j}^{2}+\mu^{2}\right)\left(\omega_{j}^{2}+\nu^{2}\right)}-\frac{\omega_{j}(\mu+\nu)\cos\left(t\omega_{j}\right)}{\left(\omega_{j}^{2}+\mu^{2}\right)\left(\omega_{j}^{2}+\nu^{2}\right)}\right]\\
 & b_{j1}(t{\color{black}\rightarrow}\infty)^{2}+\omega_{j}^{2}b_{j2}(t{\color{black}\rightarrow}\infty)^{2}=\frac{c_{j}^{2}}{M^{2}}\frac{1}{\text{(\ensuremath{\omega_{0}^{2}}-\ensuremath{\omega_{j}^{2})^{2}}+\ensuremath{\eta^{2}\omega_{j}^{2}}}}
\end{align*}

where we can confirm that $b_{j1}(0)=b_{j1}(0)=\dot{b}_{j1}(0)=\dot{b}_{j1}(0)=0$.

\section*{Appendix II}

Denote the eigenvector of $Q(t)=a_{1}(t)Q_{0}+a_{2}(t)\dot{Q}_{0}$
with eigenvalue $Q_{1}$ as $u_{Q_{1}}(Q_{0},t)$, where $Q_{0}\equiv q_{0}-\sigma_{z}d,\dot{Q}_{0}\equiv-\frac{i\hbar}{M}\frac{\partial}{\partial q_{0}}\equiv-\frac{i\hbar}{M}\frac{\partial}{\partial Q_{0}}$,
$Q(t)$ and $\dot{Q}(t)$ are understood as a function of $q_{0}-\sigma_{z}d,\dot{q}_{0}$
from Eq.(\ref{Qcd}). Since $\sigma_{z}$ commutes with $Q_{0},\dot{Q}_{0}$,
there is no ambiguity for the functions of $Q_{0},\dot{Q}_{0}$ regarding
the order of $\sigma_{z}$ in any product in the function. The eigenequation

\begin{align*}
 & \left(a_{1}(t)Q_{0}-\frac{i\hbar a_{2}}{M}\frac{\partial}{\partial Q_{0}}\right)u_{Q_{1}}=Q_{1}u_{Q_{1}}
\end{align*}

leads to

\[
u_{Q_{1}}=C(Q_{1},t)\exp\left[-i\frac{M}{2\hbar a_{2}}\left(a_{1}Q_{0}^{2}-2Q_{1}Q_{0}\right)\right]
\]

$C(Q_{1},t)$ is determined by orthonormal condition on $u_{Q}$,
i.e., $\int dQ_{0}u_{Q_{2}}^{*}(Q_{0})u_{Q_{1}}(Q_{0})=\delta(Q_{1}-Q_{2})$
as $C(Q_{1},t)=\left(\frac{M}{2\pi\hbar a_{2}}\right)^{\frac{1}{2}}e^{-i\frac{M}{2\pi\hbar a_{2}}\phi(Q_{1},t)}$in
Eq.(\ref{uQ}).

\section*{Appendix III Wave function probability density with Brownian motion}

We abbreviate the wave function density as $f(\sum_{j}\xi_{j1})=|\psi(q_{1}-s_{1}d-\sum_{j}\xi_{j1},s_{1},t)|^{2}$,
then using the Taylor expansion of $f$ it is straight forward to
show that

\begin{align}
 & \int\int..\int\int f(\sum_{j}\xi_{j})\prod_{j}|\chi_{j}(\xi_{j},t)|^{2}d\xi_{j}\nonumber \\
 & =\frac{1}{\prod_{j}\sqrt{2\pi}\sigma_{\xi_{j}}}\int\int..\int\int f(\sum_{j}\xi_{j})\exp\left(-\sum_{j}\frac{\xi_{j}^{2}}{2\sigma_{\xi_{j}}^{2}}\right)\prod_{j}d\xi_{j}\label{xi_averaging-1}\\
 & =\frac{1}{\sqrt{2\pi}\sigma_{\xi}}\int f(\xi)\exp\left(-\frac{\xi^{2}}{2\sigma_{\xi}^{2}}\right)d\xi\nonumber \\
 & \sigma_{\xi}^{2}=\sum_{j}\sigma_{\xi_{j}}^{2}\nonumber 
\end{align}

Thus

\begin{align}
 & |\Psi(q,s,t)|^{2}\equiv\int\int..\int\int|\Psi(q,s,\{\xi_{j}\},t)|^{2}\prod_{j}d\xi_{j}\nonumber \\
 & =\int\int..\int\int|\psi(q-sd-\sum_{j}\xi_{j},s,t)|^{2}\prod_{j}|\chi_{j}(\xi_{j},t)|^{2}d\xi_{j}\label{wave_density}\\
 & =\frac{1}{\sqrt{2\pi}\sigma_{\xi}}\int|\psi(q-sd-\xi,s,t)|^{2}\exp\left(-\frac{\xi^{2}}{2\sigma_{\xi}^{2}}\right)d\xi\nonumber 
\end{align}

For probability density of Eq.(\ref{psiplus}),with $|\psi(Q,t)|^{2}=|\psi(q-d,t)|^{2}$
in Eq.(\ref{psiQ}) we have

\begin{align*}
 & |\Psi_{+}(q,t)|^{2}=\cos^{2}(\theta/2)\frac{1}{\sqrt{2\pi}\sigma_{\xi}}\int|\psi(q-d-\xi,1,t)|^{2}\exp\left(-\frac{\xi^{2}}{2\sigma_{\xi}^{2}}\right)d\xi\\
 & =\frac{1}{\sqrt{2\pi}\sigma_{\xi}}\int\left(\frac{1}{2\pi\sigma_{Q}^{2}}\right)^{\frac{1}{2}}\exp\left(-\frac{\left(q-sd-\xi-a_{1}z\right)^{2}}{2\sigma_{Q}^{2}}\right)\exp\left(-\frac{\xi^{2}}{2\sigma_{\xi}^{2}}\right)d\xi\\
 & =\cos^{2}(\theta/2)\frac{1}{\sqrt{2\pi}}\frac{1}{\sigma_{Q\xi}}\exp\left(-\frac{\left(q-d+a_{1}d\right)^{2}}{2\sigma_{Q\xi}^{2}}\right)
\end{align*}

where the width of the distribution is $\sigma_{Q\xi}^{2}=\sigma_{Q}^{2}+\sigma_{\xi}^{2}$.
The same way the probability of the lower potential well with spin
down is

\[
|\Psi_{-}(q,t)|^{2}=\sin^{2}(\theta/2)\frac{1}{\sqrt{2\pi}}\frac{1}{\sigma_{Q\xi}}\exp\left(-\frac{\left(q+d-a_{1}d\right)^{2}}{2\sigma_{Q\xi}^{2}}\right)
\]

\end{document}